\documentclass[11pt]{article}
\usepackage[T1]{fontenc}

\makeatletter

\providecommand{\LyX}{L\kern-.1667em\lower.25em\hbox{Y}\kern-.125emX\@}

\usepackage[T1]{fontenc}
\usepackage{times}
\usepackage{geometry}
\usepackage{graphics}
\usepackage{setspace}


\begin{document}


\textbf{~}

\textbf{Optical modulation at around 1550 nm in a InGaAlAs optical waveguide
containing a InGaAs/AlAs resonant tunneling diode}

~\\
 J. M. L. Figueiredo a), A. R. Boyd, C. R. Stanley, and C. N. Ironside \\
 \emph{Department of Electronics and Electrical Engineering, University of Glasgow,
Glasgow G12 8LT, United Kingdom}\\
\\
 S. G. McMeekin \\
 \emph{Cardiff School of Engineering, University of Wales Cardiff, PO Box 917,
Newport Rd., Cardiff NP2 1XH, United Kingdom}\\
\\
 A.M. P. Leite \\
 \emph{Centro de Física do Porto - ADFCUP, Universidade do Porto, Rua do Campo
Alegre 687, 4169-007 PORTO, Portugal}

~

We report electro-absorption modulation of light at around 1550 nm in a unipolar
InGaAlAs optical waveguide containing a InGaAs/AlAs double-barrier resonant
tunneling diode (DB-RTD). The RTD peak-to-valley transition increases the electric
field across the waveguide, which shifts the core material absorption band-edge
to longer wavelengths via the Franz-Keldysh effect, thus changing the light-guiding
characteristics of the waveguide. Low-frequency characterisation of a device
shows modulation up to 28 dB at 1565 nm. When dc biased close to the negative
differential conductance (NDC) region, the RTD optical waveguide behaves as
an electro-absorption modulator integrated with a wide bandwidth electrical
amplifier, offering a potential advantage over conventional pn modulators.

~

~

a) Also with the Centro de Física do Porto - ADFCUP, Universidade do Porto,
Rua do Campo Alegre 687, 4169-007 PORTO, Portugal. Electronic mail: jlfiguei@fc.up.pt

\newpage

Because of their intrinsic high-speed response and potential for electrical
gain over a wide bandwidth, resonant tunneling diodes (RTDs) have been proposed
by several groups\( ^{1-3} \) for optoelectronic applications. Previously,
we reported work on a GaAs/AlAs RTD that was sucessfully integrated in a unipolar
GaAs-AlGaAs optical waveguide,\( ^{4} \) and high-speed optical modulation
(up to 18 dB) combined with electrical gain was demonstrated. This device operated
around 900 nm.\( ^{5} \) For devices functionning at the usual optical communication
wavelengths, 1300 nm or 1550 nm, applications could include, for example, optical
distribution of modulated millimeter-wave frequency carriers for mobile communication
systems. In this paper we describe a resonant tunneling diode electro-absorption
modulator (RTD-EAM) operating at wavelengths around 1550 nm.

The operation of the device is based on a RTD within an optical waveguide which
introduces a non-uniform electric field distribution across the waveguide core.
The electric field becomes strongly dependent on the bias voltage, due to accumulation
and depletion of electrons in the emitter and collector sides of the RTD, respectively.
Depending on the dc bias operating point, a small high frequency ac signal (<1
V) can induce high-speed switching. This produces substantial high-speed modulation
of the waveguide optical absorption coefficient at a given wavelength near the
material band-edge via the Franz-Keldysh effect\( ^{4} \) and, therefore, modulates
light at photon energies lower than the waveguide core band-gap energy. The
modulation depth can be considerable because, under certain conditions, the
RTD operation point switches well into the two positive differential resistance
portions of the current-voltage I-V characteristic, with a substantial part
of the terminal voltage dropped across the depleted region in the collector
side.\( ^{5,6} \) The advantage of the RTD-EAM compared to conventional pn
modulators is that, when dc biased close to the negative differential conductance
(NDC) region, the device behaves as an optical waveguide electro-absorption
modulator integrated with a wide bandwidth electrical amplifier.

The high-frequency and large modulation depth characteristics of the RTD-EAM
are a direct consequence of the carrier transport mechanisms across the RTD
and the waveguide depletion region. They are closely related to the material
system and the specific device structure. High-speed performance can be improved
by increasing the differential negative conductance, \( G_{n} \), or decreasing
the series resistance, \( R_{s} \). The velocity of the carriers, \( v \),
and hence the carriers transit time across the whole structure, are material
and structure dependent. To obtain a larger value of \( G_{n} \), it is necessary
to achieve a high peak current density, \( J_{p} \), and high peak-to-valley
current ratio (\( PVCR \)), \( J_{p}/J_{v} \).

The demonstration and development of this new modulator concept in the InGaAs-InAlAs
material system lattice matched to InP is a promising route towards high speed,
low radio frequency (rf) power consumption, optoelectronic converters (rf-optical
and optical-rf), because it can cover the wavelength range 1.0 to 1.6 \( \mu  \)m
where optical fibers have the lowest loss and chromatic dispersion. For wavelengths
around 1550 nm, we employ a unipolar In\( _{0.53} \)Ga\( _{0.42} \)Al\( _{0.05} \)As
optical waveguide containing a InGaAs/AlAs double-barrier resonant tunneling
diode (RTD). Furthermore, due to a smaller effective mass of the electrons in
InGaAs (0.045\( m_{0} \) compared to 0.067\( m_{0} \) for GaAs), and a larger
\( \Gamma _{InGaAs} \)-X\( _{AlAs} \) barrier height (0.65 eV compared to
0.20 eV for GaAs/AlAs) which will reduce the parasitic \( \Gamma  \)-X mediated
transport, the InGaAs-InAlAs material system has improved tunneling characteristics
with a superior peak-to-valley current ratio, evident in the dc current-voltage
characteristic. In addition, a specific contact resistivity below than 10\( ^{-7} \)
\( \Omega  \)cm\( ^{2} \) and a saturation velocity above 10\( ^{7} \) cm/s
can be achieved changing the material to InGaAs-InAlAs,\( ^{7,8} \) (for GaAs/AlGaAs,
typical metal to n\( ^{+} \)- GaAs contacts have a specific contact resistivity
of about 10\( ^{-6} \) \( \Omega  \)cm\( ^{2} \), and the saturation velocity
of electrons in GaAs layers is less than 10\( ^{7} \) cm/s). Because InGaAs/AlAs
RTDs can present higher peak current density and smaller valley current density,
higher-speed operation can be expected. In summary, compared to the GaAs/AlAs
system, the use of a InGaAs/AlAs RTD in a InGaAlAs optical waveguide not only
permits operation at optical communication wavelengths but also leads to a significant
improvement in the electrical characteristics of the device.

The InGaAlAs RTD optical waveguide structure was grown by Molecular Beam Epitaxy
in a Varian Gen II system, on a n\( ^{+} \) InP substrate {[}Fig. 1(a){]}.
It consists of two 2 nm thick AlAs barriers separated by a 6 nm wide InGaAs
quantum well, sandwiched between two 500 nm thick moderately doped (Si: \( 5\times 10^{16} \)
cm\( ^{-3} \)) In\( _{0.53} \)Ga\( _{0.42} \)Al\( _{0.05} \)As spacer layers
which form the waveguide core. The InP substrate and the top heavily doped (Si:
\( 2\times 10^{18} \) cm\( ^{-3} \)) InAlAs region provide the waveguide cladding
layers, which confine the light in the direction parallel to the double barrier
plane. A \( \delta  \)-doped InGaAs cap layer was provided for formation of
Au-Ge-Ni ohmic contacts. With suitable design, good overlap can be achieved
between the electric field and the modal distribution of the waveguide; the
longitudinal character of the interaction allows large light modulation to be
achieved.

Ridge waveguides (2 to 6 \( \mu  \)m wide) and large-area mesas on each side
of the ridges were fabricated by wet-etching. Ohmic contacts (100 to 400 \( \mu  \)m
long) were deposited on top of the ridges and mesas. The waveguide width and
the ohmic contact length define the device active area. A SiO\( _{2} \) layer
was deposited, and access contact windows were etched on the ridge and the mesa
electrodes {[}Fig. 1(b){]}, allowing contact to be made to high frequency bonding
pads (coplanar waveguide transmission line, CPW). After cleaving, the devices
were die bonded on packages allowing light to be coupled into the waveguide
by a microscope objective end-fire arrangement.

The dc I-V characteristics of packaged devices were measured using a HP 4145
parametric analyser and show typical RTD behaviour. From the I-V characteristic
we can estimate the electric field change across the depleted portion of the
waveguide core due to RTD peak-to-valley switching. Figure 2 shows the I-V characteristic
of a 2 \( \mu  \)m \( \times  \) 100 \( \mu  \)m active area RTD. Typical
devices have peak current density around 18 kA/cm\( ^{2} \), with a peak-to-valley
current ratio (PVCR) of 4. The difference between the valley and peak voltages,
\( \Delta V \), is around 0.8 V, and the difference between the peak and valley
current densities, \( \Delta J=J_{p}\left( 1-PVCR^{-1}\right)  \), is about
13.5 kA/cm\( ^{2} \). (Our typical GaAs/AlAs devices show a PVCR around 1.5,
\( J_{p}\simeq  \) 13 kA/cm\( ^{2} \), \( \Delta V\simeq  \) 0.4 V, and \( \Delta J\simeq  \)
5 kA/cm\( ^{2} \).)

Two important figures of merit of the modulator can be estimated from the RTD
dc characteristics, and for a given material system they can be tailored by
structural design. They are the modulator bandwidth, which is related to the
10\%-90\% switching time, \( t_{R} \), of the RTD between the peak and valley
points, and the modulation depth, which is related to the peak-to-valley current
ratio. The RTD switching time can be estimated from \( t_{R}=4.4\left( \Delta V/\Delta J\right) C_{v} \),\( ^{6} \)
where \( C_{v} \) is the capacitance at the valley point per unit area (\( C_{v}=\epsilon /W \),
where \( \epsilon  \) is the dielectric constant, and \( W \) is the depletion
region width). For the present devices, with \( W=0.5 \) \( \mu  \)m and \( \epsilon =13\epsilon _{0} \),
\( t_{R}\simeq  \) 6 ps. From this switching time, we can expect devices with
a bandwidth larger than 60 GHz.

Optical characterisation of the modulator employed light from a Tunics diode
laser, tuneable in the wavelength region around the absorption edge of the In\( _{0.53} \)Ga\( _{0.42} \)Al\( _{0.05} \)As
waveguide (1480-1580 nm). The laser light was coupled into the waveguide by
a microscope objective end-fire arrangement. To measure the change in the optical
absorption spectrum induced by the peak-to-valley transition, a low frequency
rf signal was injected to switch the RTD between the extremes of the NDC region,
and a photodetector was used to measure the transmitted light. The electric
field enhancement close to the collector barrier due to the peak-to-valley transition,
\( \Delta \mathcal{E}\equiv \mathcal{E}_{v}-\mathcal{E}_{p} \), with \( \mathcal{E}_{p(v)} \)
representing the electric field magnitude at the peak (valley) point, can be
estimated using\( ^{9} \)
\begin{equation}
\label{electricfield}
\Delta \mathcal{E}\cong \frac{\Delta V}{W}+\frac{W}{2\epsilon v_{sat}}\Delta J
\end{equation}
 with \( v_{sat} \) being the electron saturation velocity in the depletion
region. Taking \( \epsilon =13\epsilon _{0} \) and \( v_{sat} \)=1\( \times  \)10\( ^{7} \)
cm/s, and assuming the depletion region to be 500 nm wide, we have \( \Delta \mathcal{E}\simeq  \)
45 kV/cm (for the GaAs based device we obtained \( \Delta \mathcal{E}\simeq  \)
19 kV/cm).

Assuming \( \mathcal{E}_{v}\gg \mathcal{E}_{p} \), \( \Delta \mathcal{E}\cong \mathcal{E}_{v} \),
the shift in the InGaAlAs waveguide transmission spectrum due to electric field
enhancement, \( \mathcal{E}_{v} \), as a result of the Franz-Keldysh effect,
is given approximately by\( ^{4} \)
\begin{equation}
\label{bandedge}
\Delta \lambda _{g}\cong \frac{\lambda _{g}^{2}}{hc}\left( \frac{e^{2}h^{2}}{8\pi ^{2}m_{r}}\right) ^{\frac{1}{3}}\mathcal{E}_{v}^{\frac{2}{3}}
\end{equation}
 where \( m_{r} \) is the electron-hole system reduced effective mass, \( h \)
is the Planck's constant, \( c \) is the light velocity, \( e \) is the electron
charge, and \( \lambda _{g} \) is the wavelength corresponding to the waveguide
transmission edge at zero bias, which is around 1520 nm. Eq. \ref{bandedge}
neglects any shift due to thermal effects as a consequence of the current flow
and effects due to the applied peak voltage. The observed transmission spectrum
shift associated with the electric field change is approximately \( \lambda _{g} \)\( \simeq  \)
43 nm, which agrees reasonably well with the 50 nm value obtained using Eq.
\ref{bandedge}, where a uniform electric field approximation is assumed. (For
the GaAs based device we obtained \( \lambda _{g} \)\( \simeq  \) 9 nm from
Eq. \ref{bandedge}, and we have observed a spectrum shift around 12 nm.) This
agreement suggests that Eq. \ref{electricfield} can be used, as a first approximation,
to determine the magnitude of the electric field enhancement due to peak-to-valley
transition.

Figure 3 shows the modulation depth as a function of the wavelength for peak-to-valley
switching induced by a low-frequency (1 MHz) square wave signal with 1 V amplitude,
for a 4 \( \mu  \)m \( \times  \) 200 \( \mu  \)m active area device biased
slightly below the peak voltage. A maximum modulation depth of 28 dB was obtained
at 1565 nm, which is approximately 10 dB higher than the maximum obtained with
a GaAs/AlAs device.\( ^{5} \)

In conclusion, optical modulation up to 28 dB has been demonstrated in InGaAlAs
optical waveguides containing a InGaAs/AlAs double-barrier resonant tunneling
diode (RTD), due to peak-to-valley switching. Integration of a RTD with an optical
waveguide, which combines a wide bandwidth electrical amplifier with an electro-absorption
modulator, opens up the possibility for a variety of operation modes (such as
modulation due to self-oscillation and relaxation oscillation). Previous results
obtained with the GaAs/AlGaAs system at 900 nm were confirmed and improved,
using the InGaAlAs quaternary compound which allows operation at useful wavelengths
in the third fiber optic communication window (around 1550 nm). The device appears
to offer a promising route towards a high speed, low power optoelectronic converter
(rf-optical and optical-rf). The high-speed optoelectronic characterisation
of this new modulator is under way, together with a study of its application
as high-speed photo-detector.

~

J.M.L. Figueiredo acknowledges FCT-PRAXIS XXI - Portugal for his Ph.D. grant.

~

\newpage

L99-1865

\begin{description}
\item [References]~
\item [~]~
\end{description}
\begin{quotation}
\( ^{1} \)A. F. Lann, E. Grumann, A. Gabai, J. E. Golub, and P. England, Appl.
Phys. Lett. \textbf{62}, 13 (1993).

\( ^{2} \)S. C. Kan, P. J. Harshman, K. Y. Lau, Y. Wang, and W. I. Wang, IEEE
Photon. Technol. Lett. \textbf{8}, 641 (1996).

\( ^{3} \)T. S. Moise. Y.-C. Kao, C. L. Goldsmith, C. L. Schow, and J. C. Campbell,
IEEE Photon. Technol. Lett. \textbf{9}, 803 (1997).

\( ^{4} \)S. G. McMeekin, M. R. S. Taylor, B. Vögele, C. R. Stanley, and C.
N. Ironside, Appl. Phys. Lett. \textbf{65}, 1076 (1994).

\( ^{5} \)J. M. L. Figueiredo, C. R. Stanley, A. R. Boyd, C. N. Ironside, S.
G. McMeekin, and A. M. P. Leite, Appl. Phys. Lett. \textbf{74}, 1197 (1999)

\( ^{6} \)E. R. Brown, C. D. Parker, S. Verghese, M. W. Geis, and J. F. Harvey,
Appl. Phys. Lett. \textbf{70}, 2787 (1997); S. Verghese, C. D. Parker, and E.
R. Brown, Appl. Phys. Lett. \textbf{72}, 2550 (1998).

\( ^{7} \)T. Nittono, H. Ito, O. Nakajima, and T. Ishibashi, Japan. J. Appl.
Phys. \textbf{27}, 1718 (1988).

\( ^{8} \)A. V. Dyadchenko, and E. D. Prokhorov, Radio Eng. and Electron Phys.
\textbf{21}, 151 (1976)

\( ^{9} \)J. M. L. Figueiredo, C. N. Ironside, A. M. P. Leite and C. R. Stanley,
\textbf{5th} IEEE International Workshop on High Performance Electron Devices
for Microwave and Optoelectronic Applications (EDMO), 352 (1997).
\end{quotation}
~

\newpage

~

L99-1865

\textbf{Figure Captions}

~

~

~

FIG. 1. (a) Schematic diagram of the wafer structure. (b) The RTD optical modulator
configuration.

~

~

~

FIG. 2. Experimental I-V characteristic of a 2 \( \mu  \)m x 100 \( \mu  \)m
active area RTD optical waveguide, showing a PVCR around 7 and a peak current
density of 18 kA/cm\( ^{2} \).

~

~

~

FIG. 3. Modulation depth enhancement as a function of wavelength, induced by
the RTD peak-to-valley transition.

~

\newpage

~

L99-1865

FIG. 1 of 3., J.M.L. Figueiredo, Applied Physics Letters

~

~

~

~

\vspace{0.3cm} \par\centering \includegraphics{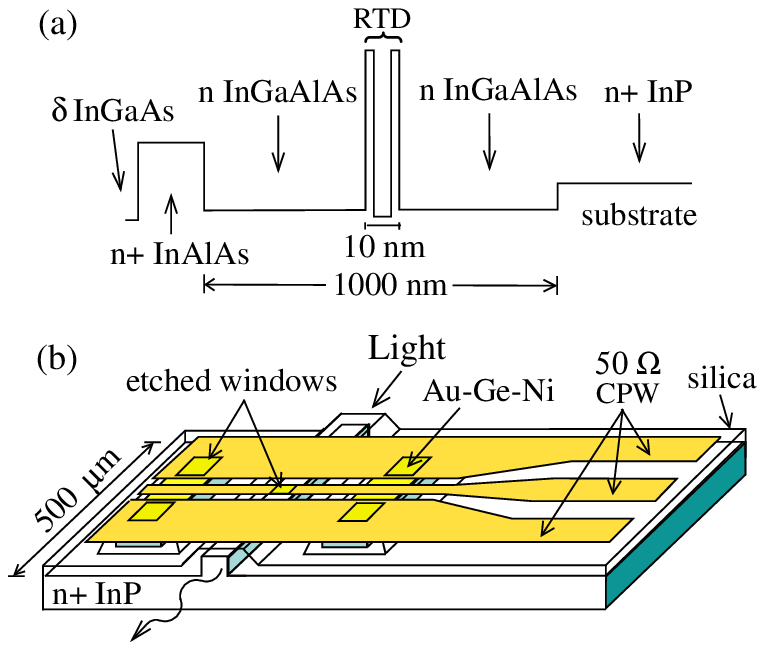} \par{} \vspace{0.3cm}

~

\newpage

~

L99-1865

FIG. 2 of 3., J.M.L. Figueiredo, Applied Physics Letters

~

~

~

~

~

~

~

\vspace{0.3cm} \par\centering \includegraphics{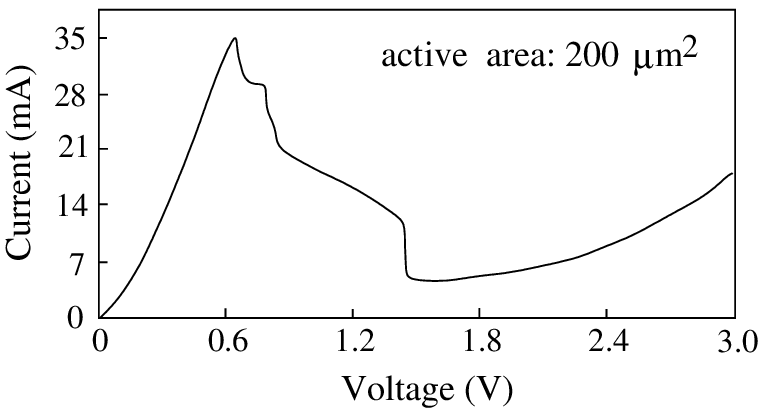} \par{} \vspace{0.3cm}

~

\newpage

~

L99-1865

FIG. 3 of 3., J.M.L. Figueiredo, Applied Physics Letters

~

~

~

~

~

~

~

\vspace{0.3cm} \par\centering \includegraphics{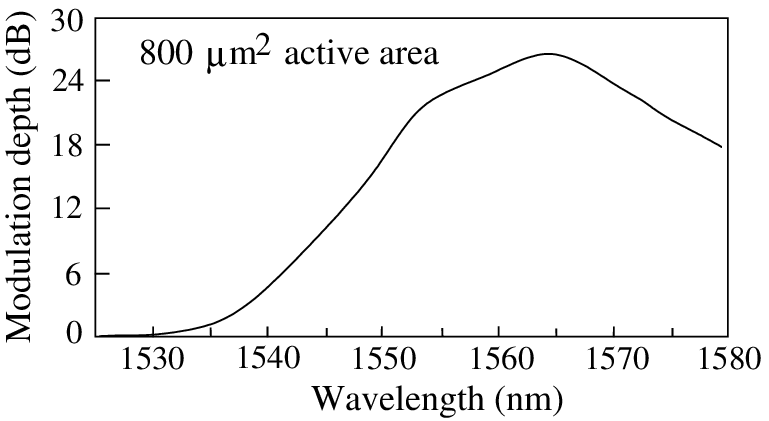} \par{} \vspace{0.3cm}

~

\end{document}